%% file: ControllableKNN.tex
\documentclass{llncs}
\usepackage{amsfonts}
\usepackage{mathtools}
\usepackage{amsmath}
\usepackage{wrapfig}
\usepackage{blindtext}
\usepackage{authblk}
\def\mathbi#1{\textbf{\em #1}}

\begin{document}
\title{Secure k-NN as a Service Over Encrypted Data in Multi-User Setting}
\author{Gagandeep Singh}
\author{Akshar Kaul}
\author{Sameep Mehta}
\institute{IBM Research India}
\affil[]{\textit{\{gagandeep\_singh,akshar.kaul,sameepmehta\}}@in.ibm.com}
\maketitle
\begin{abstract}
To securely leverage the advantages of Cloud Computing, recently a lot of research has happened in the area of \emph{``Secure Query Processing over Encrypted Data"}.
As a concrete use case, many encryption schemes have been proposed for securely processing k Nearest Neighbors (SkNN)  over encrypted data in the outsourced setting. 
Recently Zhu et al.\cite{zhu1} proposed a SkNN solution which claimed to satisfy following four properties: \textit{(1)Data Privacy, (2)Key Confidentiality, (3)Query Privacy,} and \textit{(4)Query Controllability}. 
However, in this paper, we present an attack which breaks the Query Controllability claim of their scheme.
%However, on the detailed analysis, we found that Query Controllability claim of the scheme can be broken. 
Further, we propose a new SkNN solution which satisfies all the four existing properties along with an additional essential property of \emph{Query Check Verification}. We analyze the security of our proposed scheme and present the detailed experimental results to showcase the efficiency in real world scenario.
\end{abstract}
%\begin{keywords}
%	Security, Privacy, k Nearest Neighbors, Cloud Computing, Verifiability
%\end{keywords}

\input{Introduction}
\input{PFramework}
\input{Background}

\input{Attack}
\input{kNNScheme}
\input{security}
\input{Experiments}
\input{RelatedWork}
\input{Conclusion}

\bibliographystyle{splncs03}
\bibliography{sigproc} 

\end{document}

%% file: Introduction.tex
\section{Introduction}
In recent years, cloud computing has become very popular. Many companies like Google, Amazon, and IBM etc. has started providing computing resources as a service.
% under pay-as-per-usage policy. In this setting, the enterprises outsource their data to cloud which provides the infrastructure and resources for storing and managing the data. All the subsequent operations like insertion, updates, selection and deletion are handled by cloud. 
Migration to cloud is a very lucrative option, as it provides the advantages of pay-as-per-usage, ease of scalability etc. However, the major challenge lies in the data security where enterprises have to trust the cloud with their data.
%However security of outsourced data is the biggest roadblock in adoption of clouds especially by major enterprises for their day to day usage.

%For data security, the basic solution is to encrypt data before sending it to cloud. The choice of encryption scheme for encrypting the data plays the direct role in query execution capabilities of the cloud. For example with secure block cipher encryption scheme like Advanced Encryption Standard (AES)\cite{aes} only equality predicate queries can be performed limiting the cloud only to a storage repository. With Fully Homomorphic Encryption (FHE)\cite{FHE}  scheme one can perform the complex queries involving high degree polynomial computations but the execution time is orders of time slower w.r.t. plain text execution. So the challenge here lies in building an efficient servicing solution over cloud platform that guarantees data security.
%When a data owner uses cloud services, he is implicitly trusting the cloud with all his data. This is
Trusting cloud with data is very unnerving for many data owners. To secure the data a basic solution is to encrypt it using the standard cryptographic technique such as Advanced Encryption Standard (AES)\cite{aes}. However, this reduces the cloud to a mere storage repository since AES prevents any operations on encrypted data. At the other end, Data owner can use recently developed Fully Homomorphic Encryption (FHE)\cite{FHE} to encrypt his data. FHE allows computation of any complex operations directly over the encrypted data in the cloud.
%FHE allows any complex operations to be carried out at cloud directly over encrypted data.
However, the current state of the art FHE schemes are impractical for the use in real world scenarios.

An alternate approach is to build an efficient and secure solution for a particular problem at hand. Specifically for this paper, we consider the problem of SkNN i.e. securely computing the k nearest neighbors of a query point over the encrypted data. Varying class of solution schemes has been proposed around the SkNN problem, like schemes built by using matrix multiplication based data transformation\cite{wongetal,zhu1,zhu2}, schemes that compute SkNN with some level of approximation\cite{revisited,lei2017} and the scheme built by using secure 2-part protocols in federated cloud model\cite{yousefetal} etc.

%Wong et al.\cite{wongetal} presented the Asymmetric Scalar Product Preserving Encryption (ASPE) which uses matrix multiplication based approach for transformation of data points and query points to provide security. It allows comparison of distances between a pair of data points and a query point, without revealing the distances. This property is extended to compute SkNN. 
%Yao et al.\cite{revisited} present a solution for computing approximate SkNN over 2-dimensional data points using location based partitioning of data.
%Elmehdwi at al.\cite{yousefetal} present a SkNN solution in a federated cloud model setting, which is quite different from our cloud model. Their model requires two non-colluding clouds, one of which holds the encryption key.

Most recently, Zhu et al.\cite{zhu1} presented a SkNN scheme which uses matrix multiplication based operation for secure data transformation. They treat Data Owner (DO) and Query User (QU) as the separate entities. They claim to provide following properties : (1) \textit{Data Security:} DO's data is not revealed to anyone. (2) \textit{Key Confidentiality:} DO's key is not shared with anyone. (3) \textit{Query Privacy:} The query information is known only to QU. (4) \textit{Query Controllability:} QU cannot compute encryption of a new query point without engaging in a protocol with DO.
However, in this paper we present an attack which allows QU to compute encryption of a new query point without engaging with the DO, thus breaking the Query Controllability claim of their scheme.

Further to the above mentioned four properties, we present another essential property of \emph{Query Check Verification} which a SkNN solution should provide. This property allows CS to verify whether an encrypted query being submitted by QU is actually encrypted by DO or not. In the absence of this property, QU can choose random values as encrypted query and send it to CS who treats it as a valid query. This can lead to leakage of data points to QU when he gets SkNN result set of this fake query. It also leads to wastage of CS computational resources on computing SkNN for random queries.

In this paper, we present a SkNN solution which provides all of these five desirable properties. It is the first SkNN solution to have all of these five properties. Our scheme uses matrix perturbation to maintain data security. We use an additive homomorphic encryption scheme such as Paillier cryptosystem\cite{paillier} to achieve Query Privacy. 
In our scheme, the DO and CS share a secret which is used by DO while encrypting the query points. This allows CS to verify the encrypted query being submitted by a QU is in fact encrypted by DO. We present a detailed security analysis of our scheme to show that it provides all the claimed properties. We also show a detailed empirical evaluation of our scheme to showcase that it can be used in real world scenarios.

%% file: PFramework.tex
\section{Problem Framework}
\label{ProblemFramework}
In this section, we formulate the SkNN problem in multi user setting. We describe the entities involved in the system, the desired essential properties and the adversarial model against which the solution is assumed to be secure. 
\subsection{System Entities}
%Following we describe the various entities involved in our system setting. They are shown in Fig.\ref{SystemEntityFig}.
%Various entities involved in the sKNN setting are as follows:
%Our system considers following entities as shown in Fig.\ref{SystemEntityFig} :
%\begin{wrapfigure}{r}{0.5\textwidth}
%	\begin{center}
%		\includegraphics[width=.48\textwidth]{SystemEntity.png}
%	\end{center}
%	\label{SystemEntityFig}
%	%\centering
%	\caption{System Entities}
%\end{wrapfigure}
%
\begin{figure}[h]
	\includegraphics[width=\columnwidth,height=5cm,keepaspectratio]{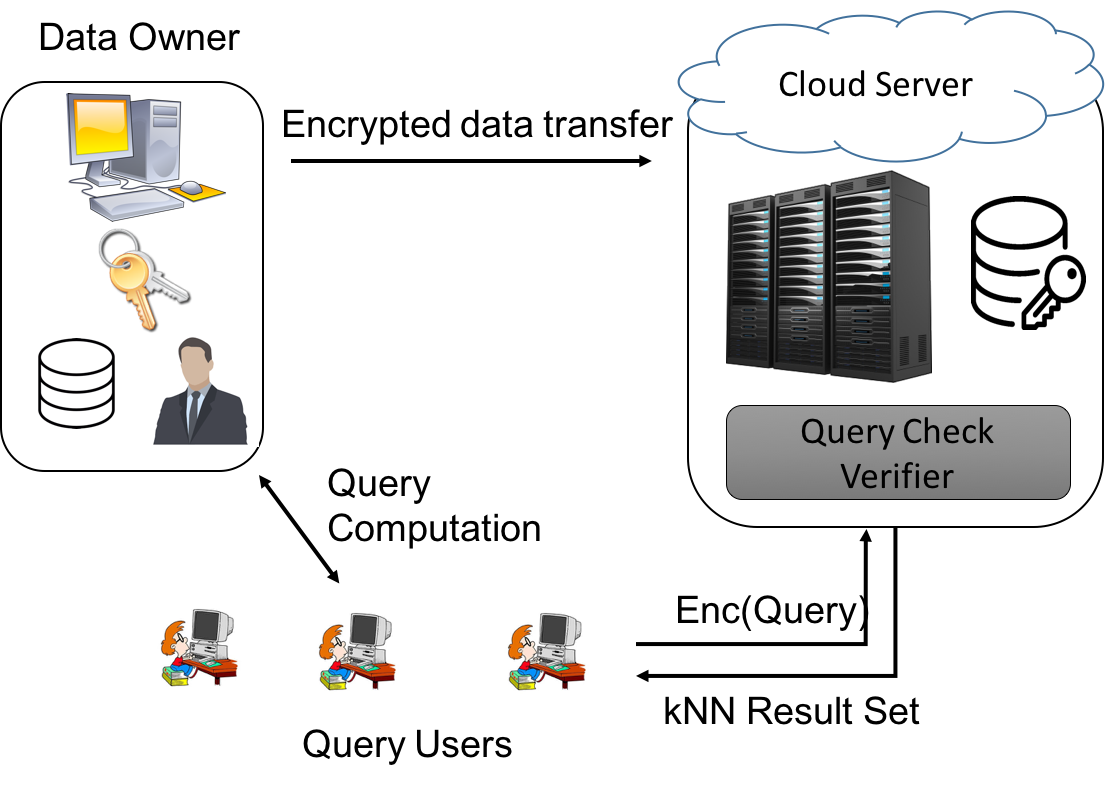}
	\centering
	\caption{System Entities}
	\label{SystemEntityFig}
\end{figure}
We consider a system containing following entities as shown in the Figure \ref{SystemEntityFig} :
\begin{enumerate}
\item \textbf{Cloud Server (CS)}: It is a third-party service provider that provides storage and computation resources to its clients. It will engage with query users for the kNN queries. Examples of CS are IBM Cloudant, Amazon AWS etc.
\item \textbf{Data Owner (DO)}: DO represents an entity who is having the propriety ownership of the data. In our setting, DO is having the data table $D$ of $d$ dimensional data points, $D =\{p_1,p_2 \cdots ,p_m\}$ s.t. each dimension of the data point is a real number, i.e. $p_i \in \mathbb{R}^d$. DO encrypts the table $D$ to $D'$ \{where $p_i$ are encrypted to $p_i'$\} and then outsources the $D'$ to CS.
\item \textbf{Query User (QU)}: These are the authorized users who want to get the kNN result for their $d$ dimensional query point $q$. In our setting, QU will first engage with DO in a secure protocol to get the valid encryption $q'$ and an authorization token $T$ to run a query over CS. QU will send $\{q',T\}$ to CS to get back the list of kNN points.  
\end{enumerate}
\subsection{Essential Properties}\label{SolutionObjective}
Our objective is to design an efficient solution to the SkNN problem which achieves the following properties :
\begin{enumerate}
\item \textbf{Data Privacy:} Data is a propriety of DO and should not be revealed to the CS. %The solution should ensure efficient computation of SkNN query and data security w.r.t. CS.
\item \textbf{Key Confidentiality:} Key generated by DO to encrypt the data is kept as a secret form everyone including QU's and CS.
%From encryption phase to the kNN query phase, DO doesn't trust key with any other party.
\item \textbf{Query Privacy:} Query point is private information of QU and should not be revealed to DO or CS in plaintext format.
\item \textbf{Query Controllability:} For every new query point, QU should engage with DO to get the encryption. QU should not be able to compute the valid encryption of a new query point by using the previously encrypted queries. 
\item \textbf{Query Check Verification:} CS should be able to verify whether the received encrypted query is authorized by DO or not.
\end{enumerate}

\subsection{Adversarial Model}
%We work in the adversarial assumption similar to \cite{zhu1}. 
We consider the most widely used \textit{Honest but curious (HBC)} adversary model.
%We assume all the entities to work under assumptions of \textit{Honest but curious (HBC)} adversary model. 
In this model, an adversary is expected to run the protocols as is expected from him (honesty), but he can do extra processing to infer more information about the encrypted data (curiosity). 
%but during the process, he could work on to get more information about the encrypted data.
%In our setting of SkNN, 
We consider the adversarial setting w.r.t. both the stored encrypted data $D'$ as well as the encrypted query points $q'$.

%In case of encrypted data $D'$ stored at CS. CS will act as an HBC adversary to infer plain text information w.r.t. $D'$. Similar to \cite{zhu1} we will consider the data privacy against the following adversarial models:
For encrypted data $D'$, CS will be the HBC adversary whose goal is to infer information about the plain text of $D'$ (i.e. about $D$). Following \cite{zhu1}, we evaluate data privacy in following adversarial models:
\begin{itemize}
\item Level 1: Adversary only knows the encrypted values of the points. His goal is to use this information to infer the plain text value.
\item Level 2: Adversary has knowledge about the existence of few plain text values along with Level 1 adversary information. However he has to figure out the respective encrypted values w.r.t. plain text values.
\end{itemize}

For query privacy, both CS and DO act as the adversaries. The DO's objective is to learn about the query point during the QueryEncryption phase while CS can try to learn about the query point from the encrypted query point and SkNN computation.
%can work on during the sKNNCal phase of the kNN computation to figure out the query point.

%% file: Background.tex
\section{Background}\label{Background}
This section describes the prior SkNN solutions which are related to our SkNN scheme. It also details out the attack to break the Query Controllability property of \cite{zhu1}.
%In this section, we describe the earlier work which act as a primary for our SkNN scheme. We  also present an attack over the claim of controllability by one of these works \cite{zhu1}.
\subsection{ASPE scheme \cite{wongetal}}
Wong et al.\cite{wongetal} were the first one to present a SkNN scheme, called ASPE (Asymmetric Scalar Product Preserving Encryption), which uses matrix multiplication based data transformation. ASPE uses an invertible matrix $\mathbi{M}$ as the key to encrypt the data points and the inverse matrix $\mathbi{M}^{-1}$ is used to encrypt the query points. This encryption has the property of preserving the scalar multiplication between data  point and query point vectors. Their scheme consists of the following procedures:
\begin{itemize}
\item \textbf{KeyGen} : DO generates a $(d+1) \times (d+1)$ invertible matrix $\mathbi{M}$ as key.
\item \textbf{TupleEncryption}: DO considers a $(d+1)$ dimension data point vector $\mathit{ p = \{p_1,p_2, \cdots, p_d,  -0.5||p||^2\}}$ and transforms it to $p'$, where $\mathit{p' = p.\mathbi{M}}$.
\item \textbf{QueryEncryption} : Considering a $(d+1)$ dimensional query point $\mathit{q = \{q_1,}$ \\$\mathit{q_2,\cdots ,q_d,1\}}$ , QU generates a random number $r>0$ and encrypt $q$ to $q'$, where $q' = \mathbi{M}^{-1}.r.q^T$.
\item \textbf{SkNNCal} : CS calculates the kNN by using the euclidean distance $\mathcal{(ED)}$ metric for comparison. For any two plain text points $p_x, p_y$ and a query point $q$ encrypted to $p_x',p_y',$ and $q'$ respectively, if $\mathcal{ED}(p_x,q) < \mathcal{ED}(p_y,q)$ then $p_x'.q' - p_y'.q' > 0$ will hold \emph{true}. This property is used for comparing the distances over the encrypted points and thus finding the kNN.
\item \textbf{TupleDecryption} : It is the inverse operation on encrypted point $p'$, i.e. $p=p'\mathbi{M}^{-1}$.
\end{itemize}
Wong et al.\cite{wongetal}  went on to present an extended scheme under stronger adversarial model. However both the schemes were presented considering QU and DO as single entity sharing keys among themselves. Hence it is not able to achieve the properties of Key Confidentiality, Query Privacy, Query Controllability and Query Check Verification.
%Thus will not be achieving the solution objective as described in section \ref{SolutionObjective}. 
\subsection{Zhu et al. Scheme \cite{zhu1}} \label{Zhuetalscheme}
Zhu et al.\cite{zhu1} built the SkNN scheme by extending ASPE. They claim that the scheme achieves \textit{Data Privacy, Query Privacy, Key Confidentiality} and \textit{Query Controllability}. 
%Following we give a brief idea of their scheme and present an attack on the query controllability claim of their scheme.
The basic idea of their scheme is same as ASPE i.e. matrix multiplication is used for transformation of data points and query points . The difference from ASPE lies in: (1)The layout of the plaintext data points and query points, (2)The co-operative execution of QueryEncryption function by DO and QU, so that QU doesn't learn the DO's key $\mathbi{M}$ and DO doesn't learn the QU's query. 

%The $(d+1)$ dimensional data point $p$ and query point $q$ are represented as follow for matrix perturbation:
The data point $p$ and query point $q$  are represnted for encryption as follows:

$\mathit{\dot{p} = \pi(S_1-2p_1,S_2-2p_2,\cdots ,S_d-2p_d,S_{d+1}+||p||^2,\pmb{\tau},\mathbi{v})}$ 

$\mathit{\dot{q} = \pi(q_1,q_2,\cdots ,q_d,1,\mathbi{R}^{\{q\}},\mathbi{0}_{\{\epsilon\}})}$ 

In data point representation $\dot{p}$, $p_i$ represent the $i^{th}$ dimension of the data point $p$, vector $\mathit{\pmb{S} = \{S_1,S_2,\cdots,S_{d+1}\}}$ and $c$ dimension vector $\pmb{\tau}$ are the constant vectors which are generated randomly during the \textit{KeyGen} function call by DO, $\epsilon$ dimensional vector $\mathbi{v}$ is chosen randomly during the \textit{TupleEncryption} function call for the data point.

In query point representation $\dot{q}$, $\mathbi{R}^{\{q\}}$ is $c$ size random vector and $\mathbi{0}_{\{\epsilon\}}$ is $\epsilon$ sized zero vector.

$\pmb{\pi}$ is an instance of random permutation chosen by DO that is applied to all data and query points. Overall, $\{\mathbf{S,\pmb{\tau},M,\pmb{\pi}}\}$ is the key of DO used during the encryption of data points and query points.

For encryption, a $n \times n$ matrix $\mathbi{M}$ is used where $n=(d+1)+c+\epsilon$. While data points are encrypted by DO as $p' = \dot{p}.\mathbi{M}^{-1}$, for query point encryption DO and QU engage among themselves to cooperatively compute the encryption $q'$ using the additive homomorphic Paillier encryption scheme \cite{paillier}. The following additive homomorphic properties of the scheme are used in the cooperative computation:
\begin{enumerate}
\item $E_{pk}(m_1+m_2) = E_{pk}(m_1) \times E_{pk}(m_2)$: Given the knowledge of public key $pk$ and the cipher text of two integers $m_1$ and $m_2$, one can compute cipher text of $m_1+m_2$ by modular multiplication of $E_{pk}(m_1)$ and $E_{pk}(m_2)$.
\item $E_{pk}(f \times m) = E_{pk}(m)^{f}$ : Similarly, given the knowledge of $pk$ and the cipher text of integer $m$. For any integer $f$, one can compute the encryption of $f \times m$ by just calculating the modular $f$ exponentiation of $E_{pk}(m)$.
\end{enumerate}

The procedure for computating query encryption between DO and QU is as follows:\\
\textbf{STEP 1}: QU generates an instance of paillier encryption scheme $\{pk,sk\}$ and encrypts each dimension of query point using the public key $pk$. It then sends $pk$  and encrypted dimensions $\{\mathit{E_{pk}(q_1),E_{pk}(q_2), \cdots ,E_{pk}(q_d)}\}$ to DO.\\
\textbf{STEP 2} : DO selects a random number $\beta_q$ and random $c$ size vector $\mathbi{R}^{(q)}$. It then constructs a partial paillier encrypted vector $\bar{q}$ s.t. :
\begin{equation*}
\bar{q} =\pi(E_{pk}(q_1)^{\beta_q},E_{pk}(q_2)^{\beta_q}, \cdots ,E_{pk}(q_d)^{\beta_q} , \beta_q, \beta_q.\pmb{R}^{(q)},\pmb{0}_{\epsilon})
\end{equation*}
 Using a vector $\bar{q}$ and matrix key $\mathbi{M}$ a paillier encrypted $n$ sized vector $\mathbi{A}^{(q)}$ is computed s.t.\\
\begin{equation*}
\mathbi{A}^{(q)}_i = \prod_{l=1}^{n} E_{pk}(\phi_l),\text{where } \phi_l = 
\begin{cases}
E_{pk}(\bar{q}_l)^{\mathbi{M}[i,l]} & \text{if }\bar{q}_l\text{ is paillier encrypted}\\
E_{pk}(\bar{q}_l \times {\mathbi{M}[i,l]}) &\text{otherwise}\\
\end{cases}
\end{equation*}
DO sends vector  $\mathbi{A}^{(q)}$ back to QU.\\
\textbf{STEP 3}: QU decrypts each component of $\mathbi{A}^{(q)}$ using secret key $sk$ to get $q'$. In brief, $q'$ is the transformation of $\dot{q}$ due to matrix multiplication s.t., $q' = \beta_q.\mathbi{M}.\dot{q}^T$.\\

The distance comparison operator for kNN calculation of this scheme differs from the ASPE scheme. For example, considering two plain text points $p_x , p_y$ and a query point $q$ s.t.  if $\mathcal{ED}(p_x,q) > \mathcal{ED}(p_y,q)$ then  correspondingly in encrypted formats  $p_x'q' - p_y'.q' > 0$ will hold true.

Though the scheme achieves the claimed properties of \textit{Data Privacy, Query Privacy}, and \textit{Key confidentiality}, it fails to achieve \emph{Query Controllability}. We now present the attack to break the Query Controllability property of the scheme.

%% file: Attack.tex
\subsection{Attack on Controllability} \label{Attack}
 During the STEP 2 of QueryEncryption, it can be observed that  $\beta_q$ w.r.t. $\dot{q}$ is multiplied to every dimension by the DO. This operation is done over the paillier encrypted partial query $\bar{q}$. Then  a matrix multiplication is performed to generate $n$ size paillier encrypted vector $\mathbi{A}^{(q)}$. Here also it can be seen that every dimension is a paillier encryption of a number that is divisible by $\beta_q$. Once QU computes the $q'$ from the paillier encrypted $\mathbi{A}^{(q)}$, it can compute the GCD of all the dimensions which is bound to be a multiple of $\beta_q$. Mostly the number extracted is $\beta_q$ as otherwise the plausible reason would be that vector $\mathbf{M}.\dot{q}^T$ have some non 1 number as GCD of the dimensions. This occurrence is highly unlikely as matrix $\mathbf{M}$ is randomly generated and w.r.t. every query, $\dot{q}$ contains a randomly generated vector $R^{(q)}$. 
 QU can then reduce  the encrypted query received in form of $q' = \beta_q.\mathbi{M}.\dot{q}^T$ to the form $q'=\mathbi{M}.\dot{q}^T$ by dividing each dimension by $\beta_q$. Similarly, all subsequent encrypted queries can be reduced to the form $\mathbf{M}.\dot{q}^T$ by QU. 
 
 QU can then build a Map for the received encrypted queries, mapping plaintext queries $q_1, q_2,\cdots, q_{\$}$  to the reduced query encryption values $q_1', q_2',\cdots, q_{\$}'$ respectively. 
 %for which he engages with DO to get the encryption. The knowledge map maps the queries of the form $\mathit{q = \{q_1,q_2,\cdots ,q_d,1\}}$  to their respective reduced encryptions of the form  $q'=\mathbi{M}.\dot{q}^T$. 
 Now for a  specific query $q_z$, which is the linear combination of subset $\mathbi{S}$ of already asked queries. QU need not engage with DO to get the encryption. Instead, he can just apply the same linear  combination over the  encrypted queries of the respective queries of subset $S$. That will be the valid encryption of the query $q_z$ for the case when $\beta_q=1$.
 
 For QU to be free from DO's encryption engagement, he must initially ask for encryption of a minimum set of query points which can act as a basis for all the points in the point space. W.r.t. kNN calculation the point space is $d+1$ dimensions where $(d+1)^{th}$ dimension is equal to value $1$. In the encryption protocol, QU  sends the paillier encryption of the $d$ dimensions of the query point and DO appends $1$ to it as a $(d+1)^{th}$ dimension. One such set of query points which can act as basis point are $\mathit{\pmb{0}^d,\pmb{e}_1,\pmb{e}_2,\cdots,\pmb{e}_d}$ where $\mathit{\pmb{0}^d}$ is $d$ dimensional zero vector and $\mathit{\pmb{e}_i}$ is the $i_{th}$ row of the $d \times d$ identity matrix. With the varying linear combination of $\pmb{e}_i$ vectors, any point can be constructed in the point space. The $\mathit{\pmb{0}^d}$ point is added to adjust the $\mathit{(d+1)^{th}}$ dimension to 1.
% Given this attack, QU just needs to get encryption of $(d+1)$  vectors, for generating encryption of any query point by himself without engaging with DO.  Basis vectors are vectors which are linearly independent of each other and every point in the vector space is just a linear combination of basis vectors. So for $d+1$ vector space QU engages with DO to get encryption of queries $e_1,e_2,\cdots,e_{d+1}$ where $e_i$ is the $i_{th}$ row of the $d+1 \times d+1$ identity matrix. 
 For example in $\mathit{(2+1)d}$ point space with queries of form $\mathit{q=\{q_1,q_2,1\}}$, the basis query points are $\mathit{(0,0), (1,0)}$ and $\mathit{(0,1)}$, every other query point can be formed by linear combination of these points and the adjustment to $\mathit{1}$ in $\mathit{3^{rd}}$ dimension can be done by point $\mathit{(0,0)}$. \\
 \\
 \textbf{\textit{Example of Attack:}} We take the example presented in \cite{zhu1} to show the attack. We will first show that the $\beta_q$ is leaked to QU. Then we will show an example for computing the encryption of new query point without the involvement of DO.
 
% Since we are attacking the controllability claim, we will only be working on \textit{QueryEncryption} algorithm of the scheme. For the detail understanding of other procedures w.r.t this example we refer  \cite{zhu1} to the readers. 
 The elements involved in the \textit{QueryEncryption} function involve, DO's key  matrix $\mathbi{M}$, permutation instance $\pi$, and QU's instance of paillier scheme $(pk,sk)$. The instances of the elements  are as follows:
\[
 \mathbi{M} = \begin{bmatrix}
   6.7 & 1.2 & 2.6 & 3.3 & 5.5 \\
9.2 & 45 & 11 & 3.2 & 19\\
17 & 1.5 & 8.3 & 2.1 & 14\\
30 & 2.9 & 16 & 20 & 6.2\\
11 & 28 & 3.6 & 23 & 13\\
\end{bmatrix},
\pi = \{3,1,4,5,2\},c=1,\epsilon=1, d=2
\]\\
 \textbf{A. Revelation of $\pmb{\beta_q}$}: Consider a query point $(13, 97)$.\\\\
 \textbf{STEP 1:} QU sends $\{E_{pk}(13),E_{pk}(97)\}$ to DO.\\
 \textbf{STEP 2:} DO choses a random $c$ dimension vector $\pmb{R}^{(q)}$ let it be $\{43\}$ and a random number $\beta_q$ let it be 131. So vector
 \begin{align*} 
\bar{q} &= \pi\{E_{pk}(13)^{131},E_{pk}(97)^{131},131,5633,0\}\\
 &=  \{131,E_{pk}(13)^{131},5633,0,E_{pk}(97)^{131}\}
\end{align*}
Thereafter DO computes vector $\pmb{A}^{(q)}$. Since there are single point decimals in matrix $\pmb{M}$ scale factor of 10 is considered during the paillier operations. The elements of vector $\pmb{A}^{(q)}$ are as below:
\begin{align*} 
\pmb{A}_{1}^{(q)} &= E_{pk}(131*67) * E_{pk}(13)^{131*12}*E_{pk}(5633*26),E_{pk}(0),E_{pk}(97)^{131*55}\\
 \pmb{A}_{2}^{(q)} &= E_{pk}(131*92) * E_{pk}(13)^{131*450}*E_{pk}(5633*110),E_{pk}(0),E_{pk}(97)^{131*190}\\
 \pmb{A}_{3}^{(q)} &= E_{pk}(131*170) * E_{pk}(13)^{131*15}*E_{pk}(5633*83),E_{pk}(0),E_{pk}(97)^{131*140}\\
 \pmb{A}_{4}^{(q)} &= E_{pk}(131*300) * E_{pk}(13)^{131*29}*E_{pk}(5633*160),E_{pk}(0),E_{pk}(97)^{131*62}\\
 \pmb{A}_{5}^{(q)} &= E_{pk}(131*110) * E_{pk}(13)^{131*280}*E_{pk}(5633*36),E_{pk}(0),E_{pk}(97)^{131*130}
\end{align*}\\
\textbf{STEP 3:} QU receives $\pmb{A}^{(q)}$ from DO, decrypts it using key $sk$ and adjusts the scale factor to get $q'$.
\begin{align*}
q' &= Dec_{sk}(A^{(q)}) /10\\
    &= ( 87455.6,381236.2,229433.4,177780.1,234594.8)
\end{align*}
Clearly from the above,QU can easily compute the $\beta_q=131$ using GCD across all the dimensions of $q'$.\\
\\
\textbf{B. Encryption of new Query Point:} Assume, QU wants to compute the encryption of query point $(13,81)$ without the DO's involvement. It will begin by asking encryption of query points $q_1 = (0,0),q_2 = (0,1),q_3 = (1,0)$.\\
For $q_1$,
\begin{align*}
&q_1' = \{ 1833.3, 7354.2, 5760.3,11046.0,2574.6\} \text{ when DO selects }\beta_{q_1}=21\text{ and }\\
&R^{(q_1)}=31 \\
&q_1'/\beta_{q_1} = \{ 87.3,350.2,274.3,526.0,122.6\}\text{ when }R^{(q_1)}=31 
\end{align*}
Similarly  for $q_2$, $q_3$ we have :
\begin{align*}
q_2'/\beta_{q_2} &= \{462.0,1931.2,1466.9,2804.2, 646.8\}\text{ when }R^{(q_2)}=173 \\
q_3'/\beta_{q_3} &= \{260.1,1121.2,823.6,1584.9,388.2\}\text{ when }R^{(q_3)}=97 
\end{align*}
For  $q_{new}=(13,81)$, we can see that $q_{new} = 13*q_3 + 81*q_2-(81+13-1)q_1$. Using this linear combination directly over the encrypted query points we get:
\begin{align*}
q_{new}'/\beta_{q_{new}} &= 13*(q_3'/\beta_{q_3}) + 81*(q_2'/\beta_{q_2})-(81+13-1)(q1'/\beta_{q_1})\\
&= \{32684.4, 138434.2, 104015.8, 198825.9,  46035.6\}
\end{align*}
 The computed $q_{new}'/\beta_{q_{new}}$ is  correct  encryption of $q_{new}$. If the DO selects $R^{(q_{new})} = 12391$ and $\beta_{q_{new}}=1$, QU will get the same encrypted value w.r.t. $q_{new}$.

%% file: kNNScheme.tex
\section{SkNN Scheme}\label{kNNScheme}
In this section, we present a new SkNN solution scheme that satisfies all the properties listed in section \ref{SolutionObjective} and then present the security arguments for the security of the proposed solution.
\subsection{Verifiable SkNN Scheme}\label{verifiabilityscheme}
Following we present the formal procedures for Verifiable SkNN scheme that satisfies all the desired properties.\\\\
%Our new SkNN scheme is described by the following procedures:\\
%This scheme is built over the rectified Query Controllable scheme, to empower the CS to verify legit queries. 
%This is necessary as CS should check the validity of queries that the queries are encrypted by DO for the specific query user. In the adversarial scenario, QU can just send any n dimension random vector to CS for the sKNNCal procedure to evaluate kNN. This leads to illicit access of some random k encrypted tuples by the QU and the wastage of CS's computational resources for evaluating kNN for the fake query. 
\textbf{KeyGen:} There are two sets of keys generated during this phase: (1)Keys that are generated by DO for encrypting the data points, (2)Keys that are jointly agreed upon by DO and CS, to be used for query verification.
\begin{itemize}
	\item DO generates parameters $\mathbf{\{M,\pmb{\pi,\tau},S\}}$ as key, where $\pmb{M}$ is a randomly generated $n \times n$ invertible matrix, $\pi$ is an instance of random permutation of $n$ values, and $\tau,S$ are randomly chosen $c, (d+1)$ dimensional vectors respectively. Here, $n=(d+1)+c+\epsilon$.
	\item  DO and CS jointly agree upon key pair $<K_{SBC},\mathbf{W}>$ where $K_{SBC}$ is key of secure block cipher like AES and $\mathbf{W}$ is randomly generated $\eta \times \eta$ invertible matrix where $\eta = n + l$.\
\end{itemize} 
\textbf{TupleEncryption:} DO encrypts the d dimensional data point $p$ using the keys generated in KeyGen procedure. For encryption, DO first generates the $\epsilon$ dimensional random vector $\pmb{v}$ and then considers the vector $\dot{p}$, s.t. $\mathit{\dot{p} = \pi(S_1-2p_1,}$\\$\mathit{S_2-2p_2,\cdots,S_d-2p_d,S_{d+1}+||p||^2,\pmb{\tau},\mathbi{v})}$ and encrypts it as follows:
\begin{equation*}
p' = \dot{p}.\mathbi{M}^{-1}
\end{equation*}
\textbf{QueryEncryption:} This procedure essentially provides the security w.r.t. desired properties of \textit{Query Privacy, Key Confidentiality, Query Controllability,} and \textit{Query Check Verification}. The procedure is jointly executed by DO and QU as below:
\begin{itemize}
	\item[] \textbf{STEP 1:} QU generates an instance of paillier encryption scheme $\{pk,sk\}$ and encrypts each dimension of query point using the public key $pk$. It then sends $pk$  and encrypted dimensions $\{\mathit{E_{pk}(q_1),E_{pk}(q_2), \cdots ,E_{pk}(q_d)}\}$ to DO.
	\item[] \textbf{STEP 2:} DO selects a random number $\beta_q$ and random $c$ size vector $\mathbf{R}^{(q)}$. It then constructs a partial paillier encrypted vector $\bar{q}$ s.t.
	\begin{equation*}
	\pmb{\bar{q}} =\pi\{\mathit{E_{pk}(q_1)^{\beta_q},E_{pk}(q_2)^{\beta_q}, \cdots ,E_{pk}(q_d)^{\beta_q} , \beta_q, \mathbi{R}^{(q)},\pmb{0}_{\epsilon}}\} 
	\end{equation*}
	Using the vector $\pmb{\bar{q}}$ and matrix key $\pmb{M}$ a paillier encrypted $n$ sized vector $\pmb{A}^{(q)}$ is computed s.t.\\
	\begin{equation*}
	\mathbi{A}^{(q)}_i = \prod_{l=1}^{n} E_{pk}(\phi_l),\text{where } \phi_l = 
	\begin{cases}
	E_{pk}(\bar{q}_l)^{\mathbi{M}[i,l]} & \text{if }\bar{q}_l\text{ is paillier encrypted}\\
	E_{pk}(\bar{q}_l \times {\mathbi{M}[i,l]}) &\text{otherwise}\\
	\end{cases}
	\end{equation*}
	\item[] \textbf{STEP 3:} DO choses $l$ size query check vector $\pmb{C}^{(q)}$ and encrypts it with \emph{Secure Block Cipher} using the key $K_{SBC}$, i.e $T=E_{K_{SBC}}(C^{(q)})$. DO then appends the check vector $\pmb{C}^{(q)}$ to the end of vector $\pmb{A}^{(q)}$, i.e. $\pmb{\hat{q}} = \{\pmb{A}^{(q)},\pmb{C}^{(q)}\}$. Using a vector $\pmb{\hat{q}}$ and matrix key $\mathbf{W}$ a paillier encrypted $\eta$ sized vector $\pmb{B}^{(q)}$ is computed s.t.
	\begin{equation*}
	\pmb{B}^{(q)}_i = \prod_{l=1}^{\eta} E_{pk}(\phi_l),\text{where } \phi_l = \begin{cases}
	E_{pk}(\hat{q}_l)^{\mathbf{W}[i,l]} & \text{if }\hat{q}_l\text{ is paillier encrypted}\\
	E_{pk}(\hat{q}_l \times {\mathbf{W}[i,l]}) &\text{otherwise}\\
	\end{cases}
	\end{equation*}
	DO sends vector  $<\pmb{B}^{(q)},T>$ back to QU.
	\item [] \textbf{STEP 4:} QU decrypts each component of $\pmb{B}^{(q)}$ using secret key $sk$ to get $\pmb{\tilde{q}}$. In brief, $\tilde{q}$ is the encryption of $\dot{q}$ obtained as transformation due to matrix multiplication operations and $T$ is a  query validity tag which will be required by CS for verification of the query.
\end{itemize}
\textbf{SkNNCal:} In this procedure, CS first verifies the query and then calculates the corresponding k nearest neighbors. The steps followed by CS on receiving the query $<\tilde{q},T>$ from QU are a follows:
\begin{itemize}
	\item [] \textbf{STEP 1:} \textit{Query Check Verification}: CS applies inverse matrix multiplication operation on $\tilde{q}$ w.r.t. matrix $\mathbf{W}$ to compute the vector $<q',\pmb{C}^{(q)}>$ and then checks the \emph{Verifiability Condition}, i.e, if ($\pmb{C}^{(q)}==D_{K_{SBC}}(T)$) . If the condition is \emph{true} CS proceeds with STEP 2, otherwise reports the \emph{fake query} back to QU.
	\item[] \textbf{STEP 2:} \textit{kNN Calculation}: CS computes the kNN by using the distance comparison operator as explained in section \ref{Zhuetalscheme} and the sub vector $q'$ computed in STEP 1. Thereafter, CS sends the kNN result set to QU.
\end{itemize}
\textbf{TupleDecryption:} This procedure just applies the inverse operation over the encrypted point $p'$, i.e. $\dot{p}= p'.\pmb{M}$. From $\dot{p}$, the data point is extracted by removing the $\pmb{S}$ vector randomization and vector $\pmb{v}.$

%% file: security.tex
\subsection{Security}\label{Security}
The security of the system is considered from the data and query point of view.\\
\textbf{Data Security:} The data encryption procedure of our scheme is similar to the scheme proposed by Zhu et al.\cite{zhu1}, likewise we claim our scheme to be secure in the level-2 adversarial model. 
In brief, the security argument for data privacy lies in the usage of random vector $\mathbi{v}$ and the invertible matrix $\mathbf{M^{-1}}$. The random vector $\mathbi{v}$ used in the data point representation $\dot{p}$, provides the randomness, as it changes with every data point, and matrix $\mathbf{M^{-1}}$ transforms the point $\dot{p}$ to $p'$ using matrix multiplication.\\ \\
%In the brief, the security argument for data privacy lies in the usage of random vector $\mathbi{v}$ that changes with every data point entry and multiplication with an invertible matrix that transforms the point $\dot{p}$ to $p'$.\\ \\
\textbf{Query Security:} \textit{QueryEncryption} and \textit{SkNNCal} procedures bring security concerns w.r.t. defined properties of \textit{Query Privacy, Key Confidentiality, Query Controllability} and \textit{Query Check Verification} for the Query Users, Data Owner and Cloud Server.

\textit{Query Privacy:} With respect to DO, QU achieves the Query Privacy by using semantically secure Paillier Cryptosystem\cite{paillier} for encrypting the query parameters that are sent to DO. DO performs further operations of QueryEncryption using the additive homomorphic properties of Paillier Scheme. With respect to CS, the security argument is similar to the Data Privacy privacy argument and lies in the usage of random vector $R^{(q)}$ and multiplication with the invertible matrices $\mathbf{M}$ and $\mathbf{W}$ that are unknown to CS.

\textit{Key Confidentiality:} The keys used by DO during \emph{QueryEncryption} process are $\mathbf{M,W},\mathbf{K_{SBC}}$. For QU to infer $\mathbf{M}$ and $\mathbf{W}$, he has to build a system of equations mapping $\dot{q}$ to $q'$. But since for each query point encryption DO chooses a random $\beta_q, R^{(q)},$ and an independent vector $C^{q}$, the random variables in such system of equations increases with the addition of every new equation, thus making the system unsolvable. The secrecy of key $K_{SBC}$ of secure block cipher depends on the strength of cipher scheme. For our scheme, we used AES which is a standard thus assuring confidentiality.

\textit{Query Controllability \& Query Check Verification:} Query Controllability and Query Check Verification properties provide double security against the presumptive fake queries from QU. While Query Controllability disallows QU to infer valid encryption of any new query, Query Check Verification property disallows SkNN execution for any random vector put forward as a query point to CS. QU can break the controllability in two ways (1) With the knowledge of keys of DO for encrypting new query, that means Key Confidentiality claim fails, which is against the presented argument for \textit{Key Confidentiality}   (2) Compute the encryption of new query using previously encrypted queries, i.e. create a valid pair of $<\tilde{q_{\psi}},T_{\psi}>$ w.r.t. new query $q_{\psi}$, where $\tilde{q_{\psi}} = \{\mathbf{W}.\{q'_{\psi},C^{(q_{\psi})}\}^T\}$, $q'_{\psi}=\mathbf{M}.\{\pi \{\beta_{q_{\psi}}.(\pmb{q_{\psi}},1),\pmb{R}^{(q_{\psi})},\pmb{0}_{\epsilon}\}\}^T$, and $T_\psi=E_{K_{SBC}}(C^{(q_{\psi})})$. Following we give the counterargument against the creation of such valid pair:
\begin{enumerate}
\item Creation of valid $\tilde{q_{\psi}}$ from previous encrypted queries require some linear combination over the encrypted counterparts. We know, for each query DO chooses $\beta_q, \pmb{R}^{(q)}$ randomly and chooses $\pmb{C}^{(q)}$ independently. And for any linear combination to work, QU must either extract and remove the vectors $\pmb{R}^{(q)}, \pmb{C}^{(q)}$, or extract and remove the value $\beta_q$ from the received encrypted query $\tilde{q}$. Both of which are perfectly hidden in the unextractable form due to matrix multiplication operations of $\pmb{M}$ and $\pmb{W}$.
%Since DO chooses a random $\beta_q, R^{(q)},$ and $C^{(q)}$ for each query independently, they just add up as new variables with every new equation w.r.t. previous queries. Also, in our scheme, these variables are inextricably hidden behind the transformation due to matrix multiplication of $\mathbb{A}$ and $\mathbb{M}$ unlike \cite{zhu1}. 
\item QU can't create any tag of any check vector as he doesn't know the key $K_{SBC}$.
\end{enumerate}
Thus the scheme is secure.

%% file: Experiments.tex
\section{Experiments}\label{Experiments}

\begin{figure}[h]
	\centering
	\begin{minipage}{0.45\textwidth}
		\centering
		\includegraphics[width=0.9\textwidth]{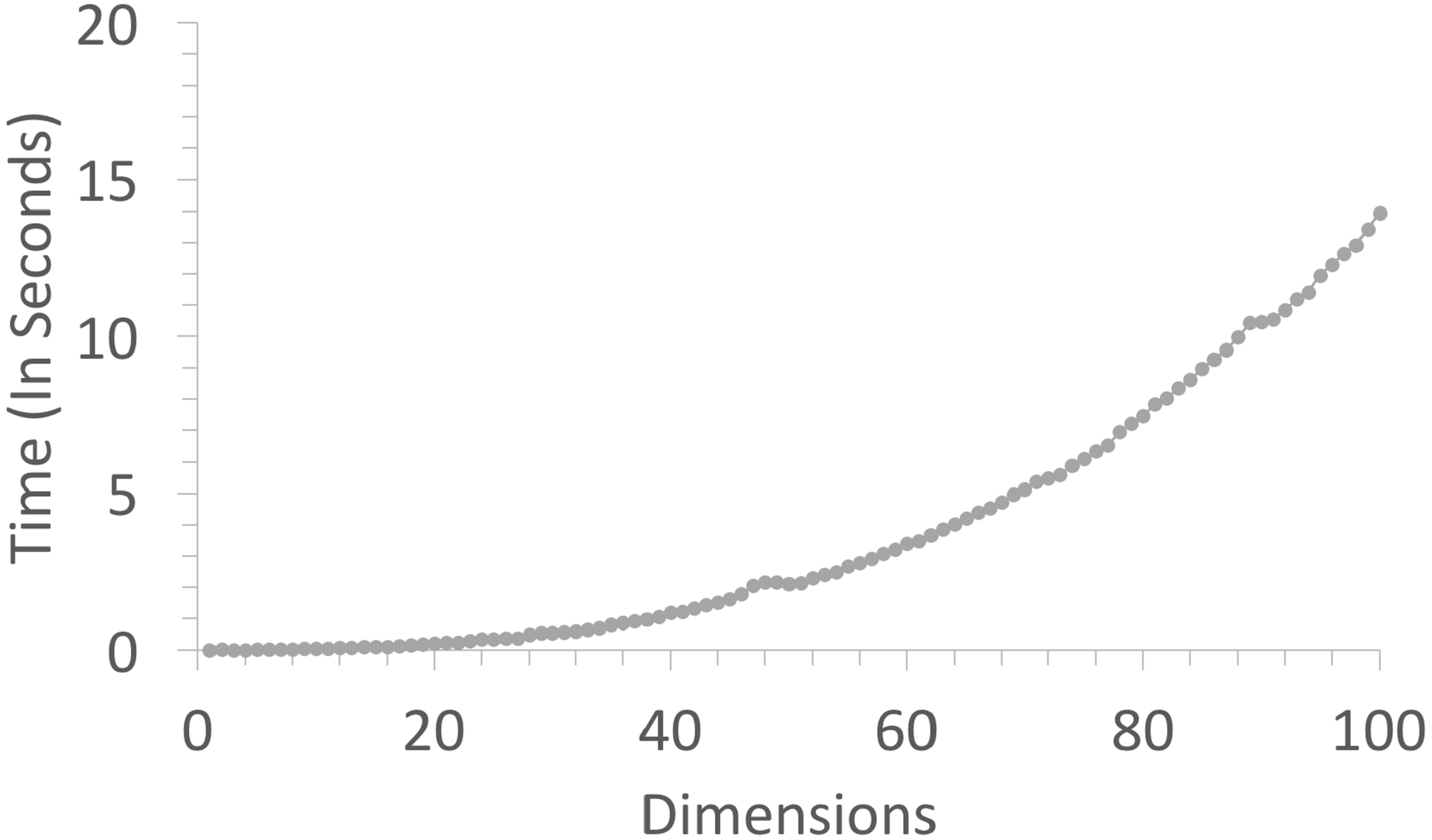}
		\caption{Time to break Zhu et al \cite{zhu1}}
			\label{fig:Attack}
	\end{minipage}\hfill
	\begin{minipage}{0.45\textwidth}
		\centering
		\includegraphics[width=0.9\textwidth]{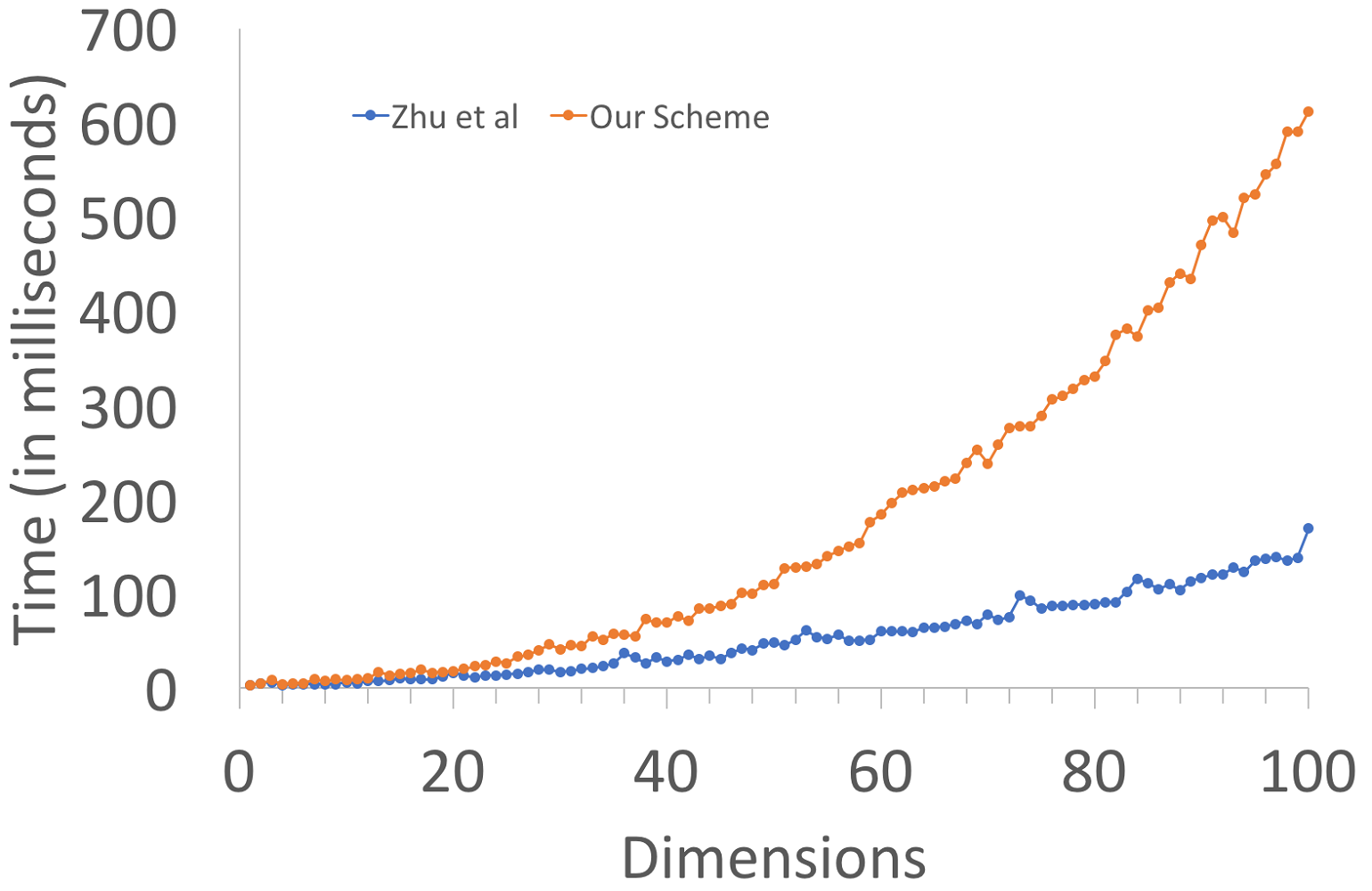}
		\caption{Query Encryption Time}
			\label{QueryEncDim}
	\end{minipage}
\end{figure}

\begin{figure}[h]
	\centering
	\begin{minipage}{0.45\textwidth}
		\centering
		\includegraphics[width=0.9\textwidth]{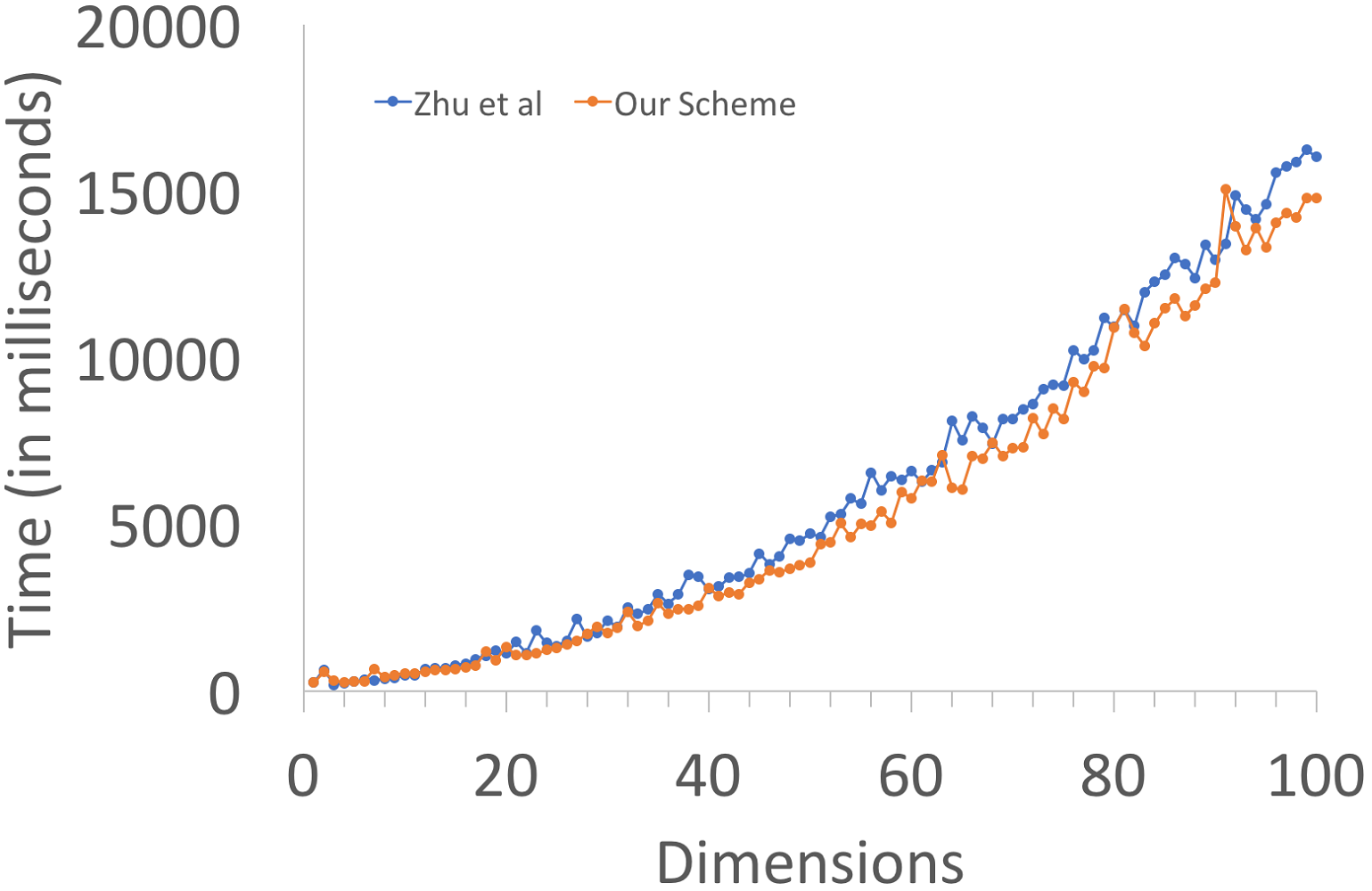}
		\caption{Database Encryption Time (Data Points = 1000000)}
		\label{DataEncDim}
	\end{minipage}\hfill
	\begin{minipage}{0.45\textwidth}
		\centering
		\includegraphics[width=0.9\textwidth]{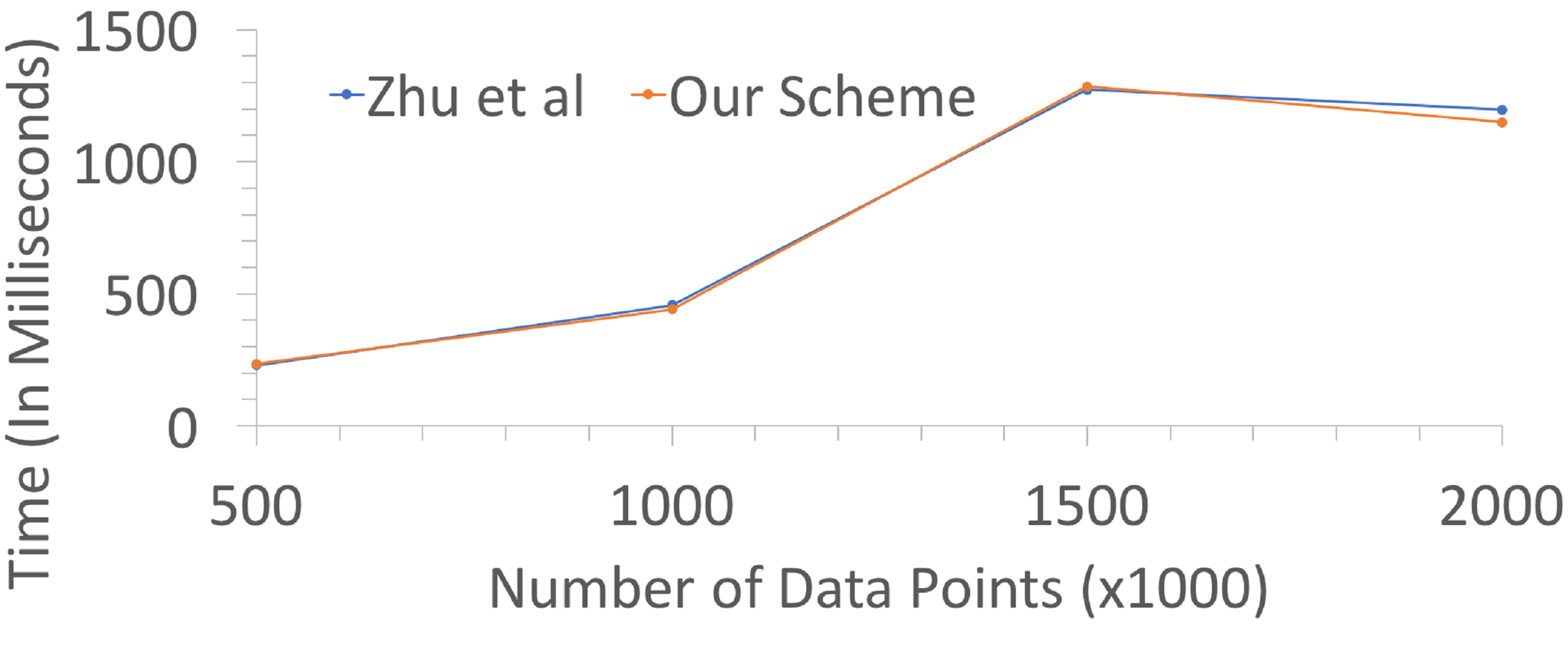}
		\caption{Database Encryption Time (Dimension = 10)}
		\label{DataEncDP}
	\end{minipage}
\end{figure}

\begin{figure}[h]
	\centering
	\begin{minipage}{0.45\textwidth}
		\centering
		\includegraphics[width=0.9\textwidth]{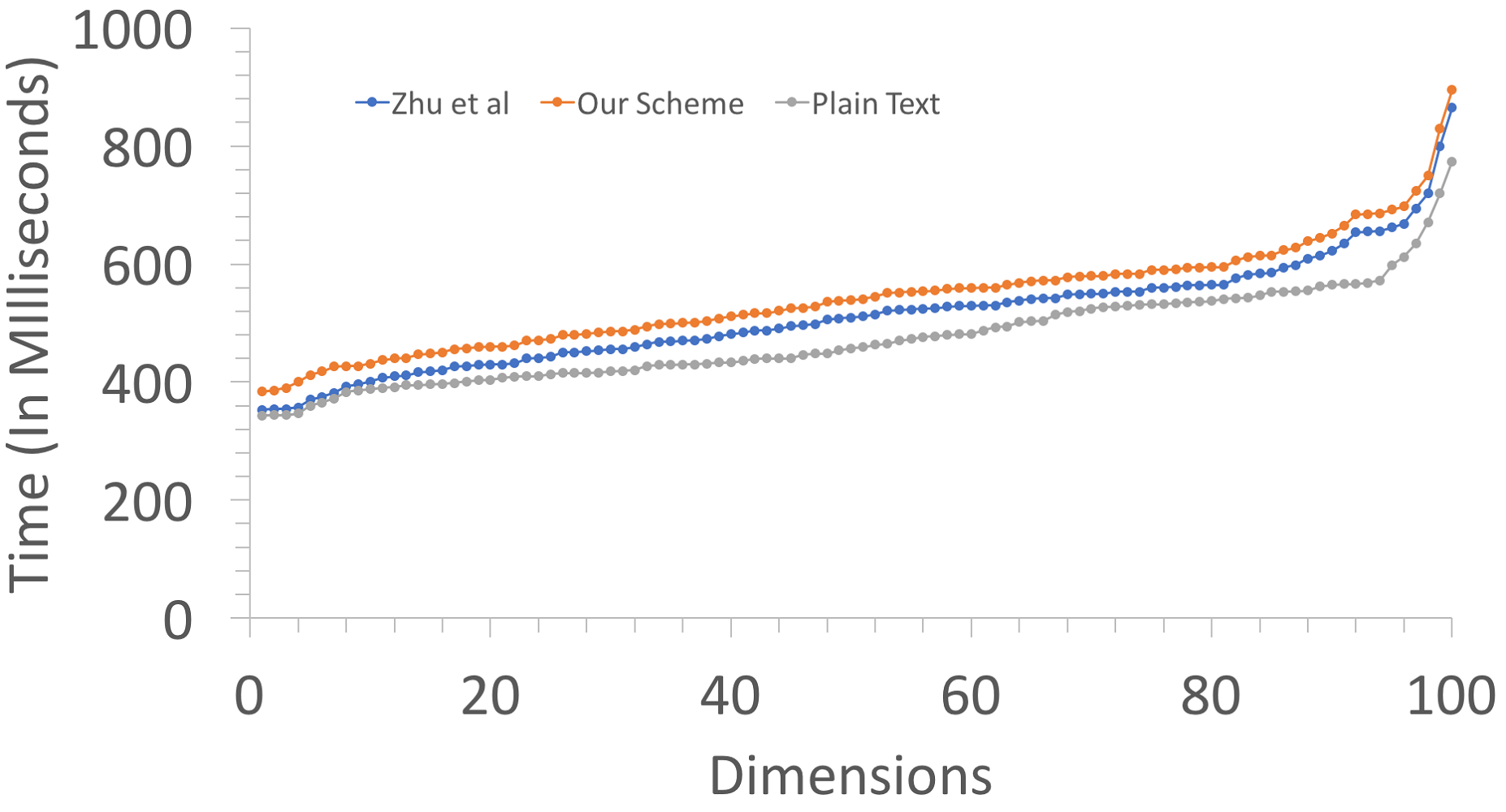}
		\caption{kNN Computation Time At Cloud (Data Points = 1000000)}
		\label{KnnDim}
	\end{minipage}\hfill
	\begin{minipage}{0.45\textwidth}
		\centering
		\includegraphics[width=0.9\textwidth]{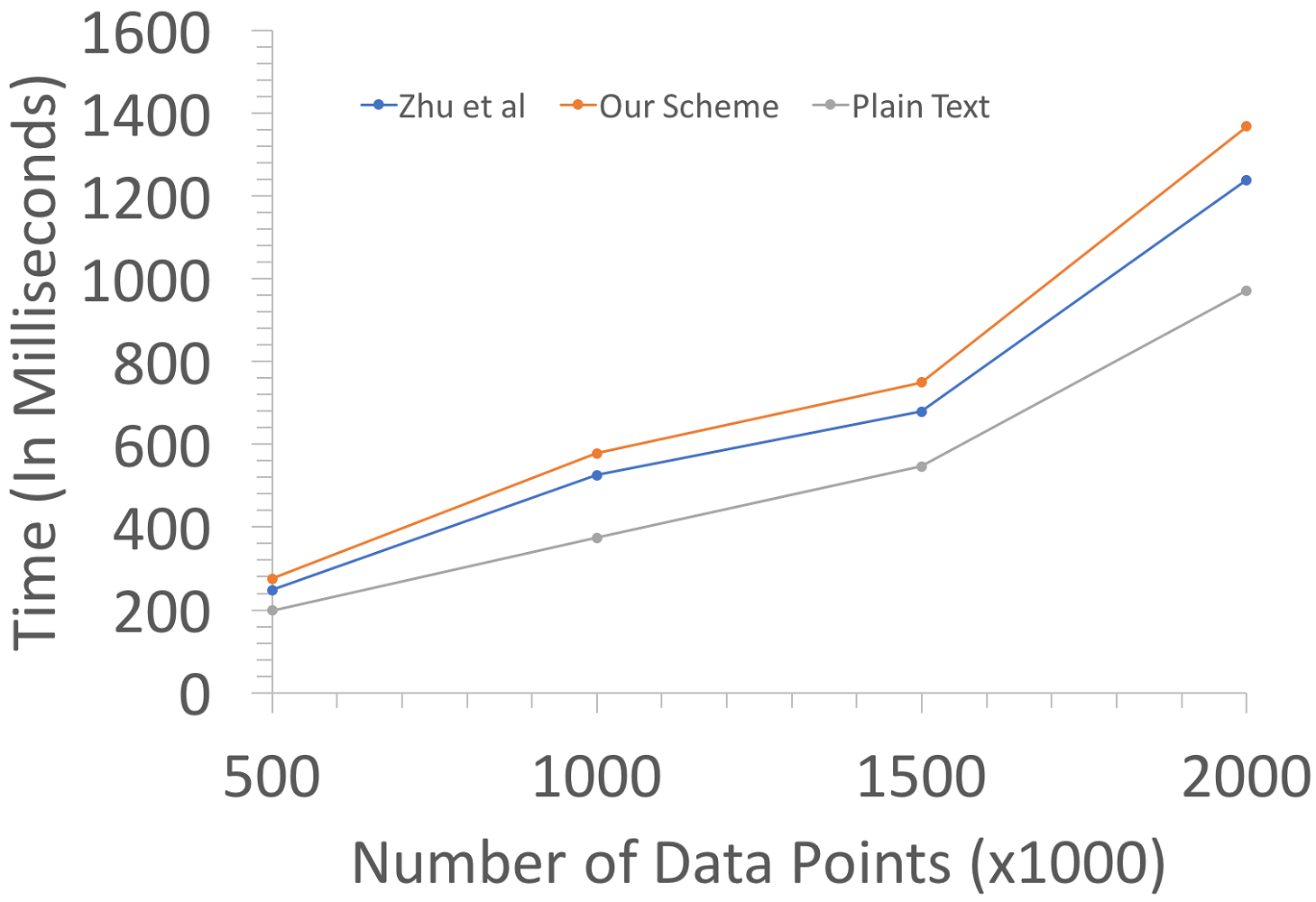}
		\caption{kNN Computation Time At Cloud (Dimension = 10)}
		\label{KnnDP}
	\end{minipage}
\end{figure}

This section presents the emperical evaluation of the SkNN scheme proposed in this paper and the controllability attack on Zhu et al.\cite{zhu1} outlined in this paper.

\textbf{Experimental Setup : }
We implemented the SkNN scheme proposed in this paper and the attack on Zhu et al.\cite{zhu1} proposed in this paper in java. We also implemented the SkNN scheme proposed by Zhu et al.\cite{zhu1} in java to compare the performance. 
We used AES encryption with a keysize of 128 bits as the symmetric cipher between the Data Owner and Cloud for query verification.
We used Paillier encryption with a keysize of 1024 bits to maintain the query privacy.
All experiments were performed on a Mac with Intel Core i7 2.7 GHz CPU and 16 GB memory. 
We set the scaling factor to $10^6$, allowing real numbers with upto 6 decimal places. Other parameters used are $c=2, \epsilon=2$ and $l=2$.

%We implement our scheme and the scheme given by Zhu et al.\cite{zhu1} using Java.  For the choice of invertible matrices key $\mathbf{M}$ and $\mathbf{W}$, the random integer matrices of appropriate sizes are chosen and then asserted for the condition $(determinant != 0)$ of invertibility, though the Java randomness almost always gives the invertible matrix at first run.   We used AES and Paillier schemes with the key sizes XXX and XXX respectively. To handle the big number arithmetic, we used \emph{BigInteger} structure of Java. All the experiments are performed on XXX with processor XXX and Ram XXX.For the experiments, we used the simulated data varying the number of dimensions from XXX to XXX, and number of records XXX to XXX. The system parameters $c, l,$ and $\epsilon$ are assigned values XXX, XXX, and XXX.

\textbf{Breaking Query Controllability of Zhu et al.\cite{zhu1} : }
We automated the attack to break the controllability of SkNN proposed by Zhu et al.\cite{zhu1}. The attack code works by first engaging with Data Owner to get the encrypted tokens for the basis vectors. Then the GCD of each of the encrypted token is calculated and then token is divided by this GCD to get reusable encrypted tokens for the basis vector.
When presented with a new random query point, the attack code uses the reusable encrypted tokens to find a encrypted token for the new query point. 
To verify whether the encrypted token generated by our attack code is correct, we run the SkNN with the encrypted token output by our attack code and compare the result with plain text kNN for the query point. 
We ran our attack code 10000 times, each time generating a new random query point, and found that the kNN returned by the encrypted token returned by our attack code and plain text kNN are same. This testifies to the correctness of our attack on the controllability of SkNN proposed by Zhu et al.\cite{zhu1}.
Figure \ref{fig:Attack} shows the time taken by our attack code to generate the encrypted token. The time shown includes the time to get the encrypted token for basis vectors. It clearly shows that our attack scales linearly with number of dimensions taking around 14 seconds for 100 dimensional data. This showcases the practicality of our attack.
 
%We tested our attack against the controllability claim of Zhu et al.\cite{zhu1}. We created the \emph{QUScript}, where for each instance of the scheme w.r.t. $d$ dimensional point size,  \emph{QUScript} first computes the encryption of query points $\mathit{0^d,e_1,e_2,\cdots,e_d}$ as described in section \ref{Attack}. From each encrypted point, factor $\pmb{\beta_q}$ is extracted out. Thereafter the encryption of 5 random query points is computed as explained in section \ref{Attack}. For correctness verification, the kNN results of the computed queries are compared with the kNN results when QU engages with DO for the same queries to compute the encrypted query. Figure XXX shows the attack results against the various instances of the scheme w.r.t. varying dimensions. The figure clearly shows almost the linear relation with changing data dimensions.

\textbf{Performance : } 
We compare the performance of Query Encryption, Database Encryption and kNN evaluation for the SkNN scheme proposed in this paper with SkNN scheme proposed Zhu et al.\cite{zhu1}.\\
Figure \ref{QueryEncDim} shows the time taken to encrypt a query point. The figure clearly shows that our query encryption runs slower than that of Zhu et al.\cite{zhu1}, though in absolute numbers there is not much difference. Even for 100 dimensional data our scheme is able to encrypt the query  in less than 1 second. Our scheme is slow because it is doing additional work of adding a l dimensional random vector to the query point along with computing AES encryption of this vector. This additional work allows us to achieve query controllability.\\
Figures \ref{DataEncDim} and \ref{DataEncDP} show the time taken to encrypt the database. Since the data encryption procedure is same in both the schemes, the time takes is also the same.\\
Figures \ref{KnnDim} and \ref{KnnDP} show the time taken to find the kNN at the cloud server. The figures clearly show that our scheme takes slightly more time than Zhu et al.\cite{zhu1}. This is due to the query verification done by the cloud server. Note that this verification is done once for query and is independent of the number of data points.\\
From the above discussion it is clear that the performance of our proposed scheme is comparable to the state-of-art scheme of Zhu et al.\cite{zhu1} and it is able to achieve query controllability  and query verification which \cite{zhu1} fails to achieve.

%We compared the \emph{QueryEncryption} and \emph{sKNNcal} procedures of our scheme with the procedures in Zhu et al.\cite{zhu1}. The data encryption and decryption procedures gave the same performance as it works the same way in both the schemes. Figure XXX shows the comparison of \emph{QueryEncryption} procedure of the two schemes. Since there is an extra step as explained in section \ref{verifiabilityscheme},  the execution times of our scheme is slightly more than \cite{zhu1}.
%Likewise in figure XXX we display the comparison of the \emph{SkNNCal} procedure with varying number of data points. Due to extra verification process at CS, our scheme takes few milliseconds more than \cite{zhu1}. 

%Overall the performance of our scheme is comparable to \cite{zhu1} and we introduce an extra feature of \emph{Query Check Verification} which helps in identification of fake queries.

%% file: RelatedWork.tex
\section{Related Work}\label{RelatedWork}
There has been a considerable amount of research in the area of secure query processing over the encrypted data. Solutions varying from specific secure predicate evaluation like text search\cite{search}, kNN\cite{yousefetal,wongetal,revisited,zhu1}, range predicate\cite{ope1,ope2,ope3,Boneh2007,Shi2007} etc. to full database level secure systems \cite{monomi,cryptdb,sdb} exist in the prior literature. For this sections we present the brief overview of the solutions build up for the secure databases and the prior secure kNN techniques. 

%\textbf{Secure Databases:} The secure database systems Monomi\cite{monomi} and Cryptdb\cite{cryptdb} used multiple encryption schemes to give support for multiple SQL predicates. Specifically, the data is encrypted multiple times by different encryption schemes. They used OPE (order preserving encryption) scheme\cite{ope2,ope3}, Paillier encryption scheme\cite{paillier} and SEARCH\cite{search} scheme to provide support for the range predicate, addition and search queries respectively over the encrypted data. Both the schemes fail to give support to complex queries where multiple operators are involved in a single query, they resort to round communication based computation between cloud and client to answer them. 
\textbf{Secure Databases:} The secure database systems Monomi\cite{monomi} and Cryptdb\cite{cryptdb} used multiple encryption schemes to give support for multiple SQL predicates. They used OPE (order preserving encryption)\cite{ope2,ope3}, Paillier encryption\cite{paillier} and SEARCH\cite{search} schemes to provide support for the range predicate, addition and search queries respectively. Both the schemes fail to support complex queries where multiple operators are involved in a single query. 
%They resort to round communication based computation between cloud and client to answer them. 
In schemes\cite{SHW3,SHW4}, cloud uses the secure hardware for secure computation where the encryption keys are stored. For complex computations, the data is first decrypted at the secure hardware, then processed for evaluation and then the answer is encrypted back and sent back to client. This model is different from our work where cloud server is not trusted with encryption keys. SDB\cite{sdb} performed query processing with a set of secure data-interoperable operators by using asymmetric secret-sharing scheme. They provided protocols to handle queries having across the column computations. 
%In schemes\cite{SHW3,SHW4} cloud use the secure hardware for secure computation over the encrypted data. The key used to encrypt the data is stored in the secure hardware. For complex computations involving high degree polynomial evaluation, the data is first decrypted at the secure hardware, then processed for evaluation and thereafter the answer is encrypted back and is sent to client. This model solution cannot be compared with our work where cloud server is not trusted with the encryption keys. SDB\cite{sdb} performed query processing with a set of secure data-interoperable operators by using asymmetric secret-sharing scheme. They provided protocols to handle queries having across the column computations. 

\textbf{SkNN Work:} \cite{ygc,sunoh,pir2} presented the schemes in different problem setting where data is stored in the plaintext format at cloud and kNN is jointly computed by cloud and query clients in oblivious manner. 
%The cloud and query client jointly compute the kNN such that cloud doesn't learn about the query and the query user doesn't learn anything about the data other than the kNN result.
We already gave an overview of the schemes of Wong et al.\cite{wongetal} and Zhu et al.\cite{zhu1,zhu2} which uses the matrix multiplication based solutions.
%Wong et al\cite{wongetal} proposed the ASPE scheme for SkNN evaluation. They were the first to use Matrix based perturbation for data point and query point transformations. To evaluate the SkNN, simple comparison is performed over the values of dot product between the transformed query point and data points. As an extended work zhu et al\cite{zhu1,zhu2} proposed a solution scheme where the encryption key is not shared with query users. And query encryption is jointly performed by data owner and query user in oblivious manner s.t. query user doesn't learn about the encryption key and data owner doesn't learn about the query. 
%Our work is mostly aligned with the works of \cite{wongetal,zhu1}.
Hu et al\cite{C5} built a system for secure index traversal using secure privacy homomorphism encryption scheme\cite{privhism}. They modeled the kNN solution by building the traversal protocols over encrypted R-Tree index structure.
% which introduces a heavy computation burden to query clients. 
Yao et al.\cite{revisited} proposed the solution in stronger security model IND-CPA\cite{criptobook} (indistinguishability under chosen plaintext attack) where the other schemes were failing. They showed that building SkNN with  IND-CPA security is at least as hard as building IND-OCPA\cite{ope2} (indistinguishability under ordered chosen plaintext attack) OPE. They presented the  IND-CPA scheme using Voronoi partitions that computes the approximate nearest neighbor for 2-dimensional data. 
Yousef et al\cite{yousefetal} presented the protocols in the federated cloud model where one cloud stores the encrypted data and the other cloud has the key information. 
%They used the paillier encryption scheme to encrypt the data. Using the homomorphic property of the scheme 
They built multiple secure protocols which are used in SkNN evaluation. Their solution provides the strong security guarantees but takes a lot of time in SkNN evaluation.
Lei et al\cite{lei2017} proposed a SkNN solution for 2-dimensional points by using LSH (Location sensitive hashing). They first construct the secure index around the data and then outsource the data and index to the cloud. Since the scheme uses LSH data structure, their result contains false positives.

%% file: Conclusion.tex
\section{Conclusion}\label{Conclusion}
In this paper we presented an attack to break the Query Controllability claim of the SkNN scheme proposed by Zhu et al. The underlying scheme revealed the extractable information in the encrypted query, which is used to build the attack. As an addendum, we introduced the essential property of Query Check Verification and presented the SkNN scheme that satisfies Query Check Verification along with all the basic properties: Query Controllability, Query Privacy, Data Privacy, and Key Confidentiality. 

We also performed the experiments to verify the attack on the scheme of Zhu et al. and to evaluate the performance of the proposed scheme.

%% file: ControllableKNN.bbl
\begin{thebibliography}{10}
\providecommand{\url}[1]{\texttt{#1}}
\providecommand{\urlprefix}{URL }

\bibitem{ope1}
Agrawal, R., Kiernan, J., Srikant, R., Xu, Y.: Order preserving encryption for
  numeric data. In: ACM SIGMOD, 2004. pp. 563--574

\bibitem{SHW3}
Arasu, A., Blanas, S., Eguro, K., Kaushik, R., Kossmann, D., Ramamurthy, R.,
  Venkatesan, R.: Orthogonal security with cipherbase. In: CIDR, 2013

\bibitem{SHW4}
Bajaj, S., Sion, R.: Trusteddb: A trusted hardware-based database with privacy
  and data confidentiality. In: IEEE TKDE, 2014. pp. 752--765

\bibitem{ope3}
Boldyreva, A., Chenette, N., Lee, Y., O’neill, A.: Order-preserving symmetric
  encryption. In: EUROCRYPT, 2009. pp. 224--241

\bibitem{ope2}
Boldyreva, A., Chenette, N., O’Neill, A.: Order-preserving encryption
  revisited: Improved security analysis and alternative solutions. In: CRYPTO,
  2011. pp. 578--595

\bibitem{Boneh2007}
Boneh, D., Waters, B.: Conjunctive, subset, and range queries on encrypted
  data. In: TCC, 2007. pp. 535--554

\bibitem{sunoh}
Choi, S., Ghinita, G., Lim, H.S., Bertino, E.: Secure knn query processing in
  untrusted cloud environments. In: IEEE TKDE, 2014. pp. 2818--2831

\bibitem{aes}
Daemen, J., Rijmen, V.: The Design of Rijndael: {AES} - The Advanced Encryption
  Standard. Information Security and Cryptography, 2002

\bibitem{privhism}
Domingo-Ferrer, J.: A provably secure additive and multiplicative privacy
  homomorphism*. In: Information Security: ISC, 2002. pp. 471--483

\bibitem{yousefetal}
Elmehdwi, Y., Samanthula, B.K., Jiang, W.: Secure k-nearest neighbor query over
  encrypted data in outsourced environments. In: IEEE ICDE, 2014. pp. 664--675

\bibitem{FHE}
Gentry, C.: Fully homomorphic encryption using ideal lattices. In: ACM STOC,
  2009. pp. 169--178

\bibitem{criptobook}
Goldreich, O.: Foundations of Cryptography: Volume 2, Basic Applications.
  Cambridge University Press, New York, NY, USA (2004)

\bibitem{C5}
Hu, H., Xu, J., Ren, C., Choi, B.: Processing private queries over untrusted
  data cloud through privacy homomorphism. In: IEEE ICDE, 2011. pp. 601--612

\bibitem{lei2017}
Lei, X., Liu, A.X., Li, R.: Secure knn queries over encrypted data:
  Dimensionality is not always a curse. In: IEEE ICDE, 2017. pp. 231--234

\bibitem{paillier}
Paillier, P.: Public-key cryptosystems based on composite degree residuosity
  classes. In: EUROCRYPT, 1999,. pp. 223--238

\bibitem{pir2}
Papadopoulos, S., Bakiras, S., Papadias, D.: Nearest neighbor search with
  strong location privacy. In: PVLDB, 2010. pp. 619--629

\bibitem{cryptdb}
Popa, R.A., Redfield, C.M.S., Zeldovich, N., Balakrishnan, H.: Cryptdb:
  Protecting confidentiality with encrypted query processing. In: ACM SOSP,
  2011. pp. 85--100

\bibitem{Shi2007}
Shi, E., Bethencourt, J., Chan, T.H.H., Song, D., Perrig, A.: Multi-dimensional
  range query over encrypted data. In: IEEE S\&P, 2007. pp. 350--364

\bibitem{search}
Song, D.X., Wagner, D., Perrig, A.: Practical techniques for searches on
  encrypted data. In: IEEE S\&P, 2000. pp. 44--55

\bibitem{ygc}
Songhori, E.M., Hussain, S.U., Sadeghi, A.R., Koushanfar, F.: Compacting
  privacy-preserving k-nearest neighbor search using logic synthesis. In: ACM
  DAC, 2015. pp. 36:1--36:6

\bibitem{monomi}
Tu, S., Kaashoek, M.F., Madden, S., Zeldovich, N.: Processing analytical
  queries over encrypted data. In: PVLDB, 2013. pp. 289--300

\bibitem{wongetal}
Wong, W.K., Cheung, D.W.l., Kao, B., Mamoulis, N.: Secure knn computation on
  encrypted databases. In: ACM SIGMOD, 2009. pp. 139--152

\bibitem{sdb}
Wong, W.K., Kao, B., Cheung, D.W.L., Li, R., Yiu, S.M.: Secure query processing
  with data interoperability in a cloud database environment. In: ACM SIGMOD,
  2014. pp. 1395--1406

\bibitem{revisited}
Yao, B., Li, F., Xiao, X.: Secure nearest neighbor revisited. In: IEEE ICDE,
  2013. pp. 733--744

\bibitem{zhu1}
Zhu, Y., Huang, Z., Takagi, T.: Secure and controllable k-nn query over
  encrypted cloud data with key confidentiality. In: JPDC, 2016. vol.~89, pp. 1
  -- 12

\bibitem{zhu2}
Zhu, Y., Xu, R., Takagi, T.: Secure k-nn computation on encrypted cloud data
  without sharing key with query users. In: Proceedings of the International
  Workshop on Security in Cloud Computing, SCC@ASIACCS, 201313,. pp. 55--60

\end{thebibliography}
